\documentclass[prl,showpacs,preprintnumbers,superscriptaddress,floatfix,twocolumn,amsmath,amssymb]{revtex4}
\usepackage{epsfig}
\usepackage{graphicx}
\usepackage{dcolumn}
\usepackage{bm}
\usepackage[active]{srcltx}
\newcommand{\be}{\begin{equation}}

\newcommand{\ee}{\end{equation}}
\newcommand{\bea}{\begin{eqnarray}}
\newcommand{\eea}{\end{eqnarray}}

\begin{document}

\preprint{}
\title{Information flow in a network model and the law of diminishing marginal returns}
\date{\today}
\author{D. Marinazzo}\affiliation{Faculty of Psychology and
Educational Sciences, Department of Data Analysis, Ghent University,
Henri Dunantlaan 1, B-9000 Ghent, Belgium\\}\author{M.
Pellicoro}\affiliation{Dipartimento di Fisica, Universit\'a degli
Studi di Bari and INFN, via Orabona 4, 70126 Bari,
Italy\\}\author{Guo-Rong Wu}\affiliation{Faculty of Psychology and
Educational Sciences, Department of Data Analysis, Ghent University,
Henri Dunantlaan 1, B-9000 Ghent, Belgium\\}\affiliation{Key
Laboratory for NeuroInformation of Ministry of Education, School of
Life Science and Technology, University of Electronic Science and
Technology of China, Chengdu 610054, China\\}\author{L.
Angelini}\affiliation{Dipartimento di Fisica, Universit\'a degli
Studi di Bari and INFN, via Orabona 4, 70126 Bari,
Italy\\}\author{S. Stramaglia}\affiliation{Dipartimento di Fisica,
Universit\'a degli Studi di Bari and INFN, via Orabona 4, 70126
Bari, Italy\\}

\date{\today}

\begin{abstract}
We analyze a simple dynamical network model which describes the
limited capacity of nodes to process the input information. For a
suitable choice of the parameters, the information flow pattern is
characterized by exponential distribution of the incoming
information and a fat-tailed distribution of the outgoing
information, as a signature of the law of diminishing marginal
returns. Similar behavior is observed in another network model,
describing in a different fashion the law of diminishing marginal
returns. The analysis of a real EEG data-set shows that similar
phenomena may be relevant for brain signals.
\end{abstract}

\maketitle

Most social, biological, and technological systems can be modeled as
complex networks, and display substantial non-trivial topological
features \cite{link,bocca}. Moreover, time series of simultaneously
recorded variables are available in many fields of science; the
inference of the underlying network structure, from these time
series, is an important problem that received great attention in the
last years. A method based on chaotic synchronization has been
proposed in \cite{yu1}, a method based on model identification has
been described in \cite{sauer}. Use of a phase slope index to detect
directionalities of interactions has been proposed in \cite{nolte}.

The inference of dynamical networks is also related to the
estimation, from data, of the flow of information between variables,
as measured by the transfer entropy \cite{schreiber,leh}. Wiener
\cite{wiener} and Granger \cite{granger} formalized the notion that,
if the prediction of one time series could be improved by
incorporating the knowledge of past values of a second one, then the
latter is said to have a {\it causal} influence on the former.
Initially developed for econometric applications, Granger causality
has gained popularity also among physicists (see, e.g.,
\cite{geweke,chen,blinoska,smirnov,dingprl,noiprl}) and eventually
became one of the methods of choice to study brain connectivity in
neuroscience \cite{bressler}. Multivariate Granger causality may be
used to infer the structure of dynamical networks from data as
described in \cite{noipre}. It has been recently shown that for
Gaussian variables Granger causality and transfer entropy are
equivalent \cite{seth}, and this framework has also been generalized
to other probability densities \cite{hlava2011}. Hence a weighted
network obtained by Granger causality analysis can be given an
interpretation in terms of flow of information between different
components of a system. This way to look at information flow is
particularly relevant for neuroscience, where it is crucial to shed
light on the communication among neuronal populations, which is the
mechanism underlying the information processing in the brain
\cite{friston}. Furthermore, recent studies have investigated the economics implications of several network types mapping brain function \cite{bull1,bull2}.

In many situations it can be expected that each node of the network
may handle a limited amount of information. This structural
constraint suggests that information flow networks should exhibit
some topological evidences of the law of diminishing marginal
returns \cite{samuelson}, a fundamental principle of economics which
states that when the amount of a variable resource is increased,
while other resources are kept fixed, the resulting change in the
output will eventually diminish \cite{lopez,lopez2}. The purpose of
this work is to introduce a simple dynamical network model where the
topology of connections, assumed to be undirected, gives rise to a
peculiar pattern of the information flow between nodes: a fat tailed
distribution of the outgoing information flows while the average
incoming information flow does not depend  on the connectivity of
the node. In the proposed model the units, at the nodes the network,
are characterized by a transfer function that allows them to process
just a limited amount of the incoming information. We show that a
similar behavior is observed also in another network model, which
describes in a different fashion the law of diminishing marginal
returns; we then show that this relevant topological feature is
found as well in real neural data.

Our model is as follows. Given an undirected network of $n$ nodes
and symmetric connectivity matrix $A_{ij}\in \{0,1\}$, to each node
we associate a real variable $x_i$ whose evolution, at discrete
times, is given by:
\begin{equation}\label{update}
x_i (t+1)=F\left( \sum_{j=1}^n A_{ij} x_j (t)\right)+\sigma
\xi_i(t),
\end{equation}
where $\xi$ are unit variance Gaussian noise terms, whose strength
is controlled by $\sigma$; $F$ is a transfer function chosen as
follows:
\begin{eqnarray}
\begin{array}{lllr}
F(\alpha)&=a\alpha& & |\alpha|<\theta\\
F(\alpha)&=a\theta& &\alpha > \theta\\
F(\alpha)&=-a\theta& &\alpha < -\theta\\
\end{array}
\label{transfer}
\end{eqnarray}
where $\theta$ is a threshold value. This transfer function is
chosen to mimic the fact that each unit is capable to handle a
limited amount of information. For large $\theta$ our model becomes
a linear map. At intermediate values of $\theta$, the nonlinearity
connected to the threshold will affect mainly the mostly connected
nodes (hubs): the input $\sum A_{ij} x_j$ to nodes with low
connectivity will remain typically sub-threshold in this case. We
consider hierarchical networks generated by preferential attachment
mechanism \cite{PRE} . From numerical simulations of eqs.
(\ref{update}), we evaluate the linear causality pattern for this
system as the threshold is varied. We verify that, in spite of the
threshold, variables are nearly Gaussian so that we may identify the
causality with the information flow between variables \cite{seth}.
We compute the incoming and outgoing information flow from and to
each node, $c_{in}$ and $c_{out}$, summing respectively all the
sources for a given target and all the targets for a given source.
Then we evaluate the standard deviation of the distributions of
$c_{in}$ and $c_{out}$, varying the realization of the preferential
attachment network and running eqs. (\ref{update}) for 10000 time
points.
\begin{figure}[ht]
\includegraphics[width=8.5cm]{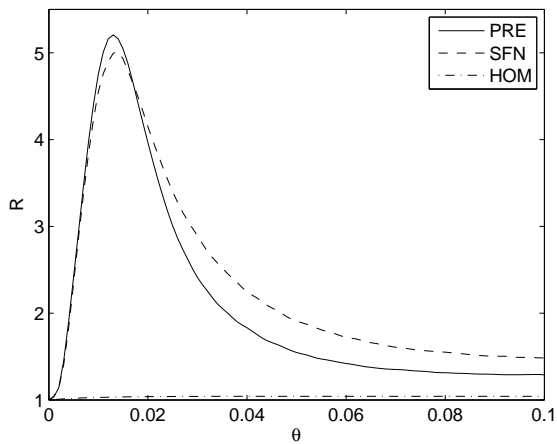}
\caption{{\rm  The ratio between the standard deviation of $c_{out}$
and those of $c_{in}$, $R$, is plotted versus $\theta$ for the three
architectures of network: preferential attachment (PRE),
deterministic scale free (SFN) and homogeneous (HOM). The parameters
of the dynamical system are $a=0.1$ and $\sigma=0.1$. Networks built
by preferential attachment are made of 30 nodes and 30 undirected
links, while the deterministic scale free network of 27 nodes is
considered. The homogeneous networks have 27 nodes, each connected
to two other randomly chosen nodes. \label{fig1}}}\end{figure}

In figure (\ref{fig1}) we depict $R$, the ratio between the standard
deviation of $c_{out}$ over those of $c_{in}$, as a function of the
$\theta$. As the  threshold is varied, we encounter a range of
for which the distribution of $c_{in}$ is much
narrower than that of $c_{out}$.  In the same figure we also
depict the corresponding curve for deterministic scale free networks
\cite{SFN}, which exhibits a similar peak, and for homogeneous
random graphs (or Erdos-Renyi networks \cite{erdos}), with $R$
always very close to one. The discrepancy between the distributions
of the incoming and outgoing causalities arises thus in hierarchical
networks. We remark that, in order to quantify the difference
between the distributions of $c_{in}$ and   $c_{out}$, here we use
the ratio of standard deviations but qualitatively similar results
would have been shown using other measures of discrepancy.
\begin{figure}[ht]
\includegraphics[width=8.5cm]{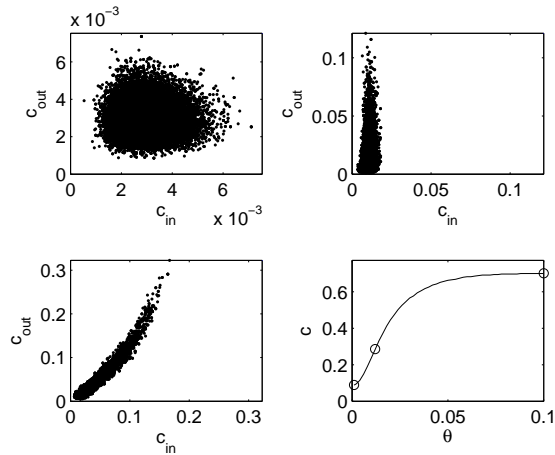}
\caption{{\rm Scatter plot in the plane $c_{in}-c_{out}$ for
undirected networks of 30 nodes and 30 links built by means of the
preferential attachment mechanism. The parameters of the dynamical
system are $a=0.1$ and $\sigma=0.1$. The points represent the nodes
of 100 realizations of preferential attachment networks, each with
10 simulations of eqs. (\ref{update}) for 10000 time points.
(Top-left) Scatter plot of the distribution for all nodes at $\theta
=0.001$. (Top-right) Contour plot of the distribution for all nodes
at $\theta =0.012$.(Bottom-left) Scatter plot of the distribution
for all nodes at $\theta =0.1$.(Bottom-right) The total causality
(obtained summing over all pairs of nodes) is plotted versus
$\theta$; circles point to the values of $\theta$ in the previous
subfigures.\label{fig2}}}\end{figure}

\begin{figure}[ht]
\includegraphics[width=8.5cm]{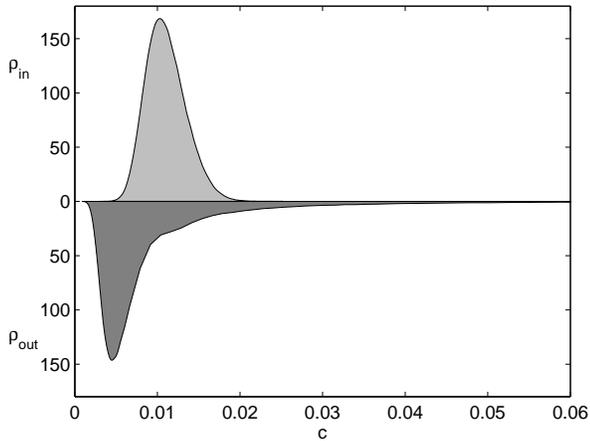}
\caption{{\rm For the preferential attachment network, at $\theta
=0.012$, the distributions (by smoothing spline estimation) of
$c_{in}$ and $c_{out}$ are depicted. Units on the vertical axis are
arbitrary. \label{fig3}}}\end{figure}

In (\ref{fig2}) we report the scatter plot in the plane
$c_{in}-c_{out}$ for preferential attachment networks and for some
values of the threshold. The distributions of $c_{in}$ and
$c_{out}$, with $\theta$ equal to 0.012 and corresponding to the
peak of fig. (\ref{fig1}), are depicted in fig. (\ref{fig3}):
$c_{in}$ appears to be exponentially distributed around a typical
value, whilst $c_{out}$ shows a fat tail. In other words,  the power
law connectivity, of the underlying network, influences just the
distribution of outgoing causalities.

\begin{figure}[ht]
\includegraphics[width=8.5cm]{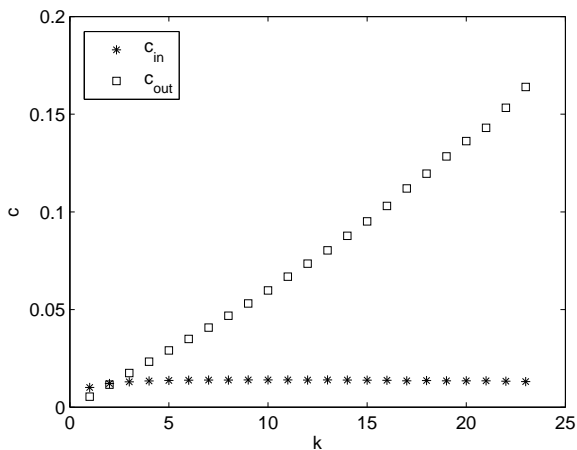}
\caption{{\rm In the ensemble of preferential attachment networks of
figure (\ref{fig2}), at $\theta =0.012$, $c_{in}$ and $c_{out}$ are
averaged over nodes with the same connectivity and plotted versus
the connectivity.\label{fig4}}}\end{figure}

In fig. (\ref{fig4}) we show the average value of $c_{in}$ and
$c_{out}$ versus the connectivity $k$ of the network node: $c_{out}$
grows uniformly with  $k$, thus confirming that its fat tail is a
consequence of the power law of the connectivity. On the contrary
$c_{in}$ appears to be almost constant: on average the nodes receive
the same amount of information, irrespectively of $k$, whilst the
outgoing information from each node depends on the number of
neighbors.

 It is worth mentioning
that since a precise estimation of the information flow is
computationally expensive, our simulations are restricted to rather
small networks; in particular the distribution of $c_{out}$ appears
to have a fat tail but, due to our limited data, we can not claim
that it corresponds to a simple power-law.


\begin{figure}[ht]
\includegraphics[width=8.5cm]{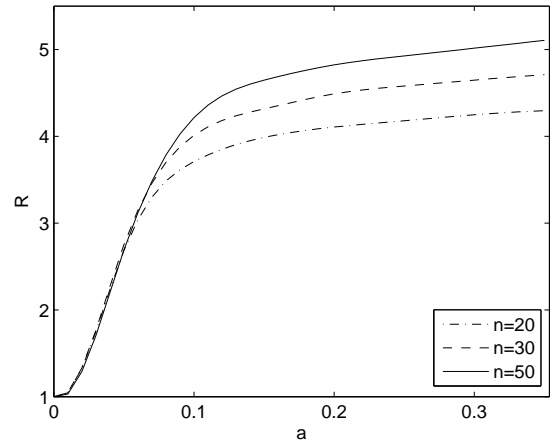}
\caption{{\rm For the model (\ref{update2}) the ratio $R$, between
the standard deviation of $c_{out}$ and those of $c_{in}$, is
depicted versus $a$. Preferential attachment networks, of $n$ nodes
and $n$ links, are considered.\label{fig5}}}\end{figure}

A fat tail in the distribution of $c_{out}$ is observed also in the
following model: to each node of an undirected network we associate
the variable $x_i$ whose evolution is
\begin{equation}\label{update2}
x_i (t+1)=a\;x_{j(t)} (t)+\sigma \xi_i(t),
\end{equation}
where $j(t)$ is a node chosen randomly, at each time $t$, in the set
of the neighboring nodes of $i$. Equations (\ref{update2})
implement, in a different way from (\ref{update}), the occurrence
that nodes may handle a limited incoming information: at each time
each node is influenced just by one other node. In figure
(\ref{fig5}) we depict $R$ as a function of $a$, for preferential
attachment networks and for different size of the networks: the
discrepancy between the distributions of $c_{in}$ and $c_{out}$
increases as the size of the network grows while keeping $a$ fixed.


\begin{figure}[ht]
\includegraphics[width=8.5cm]{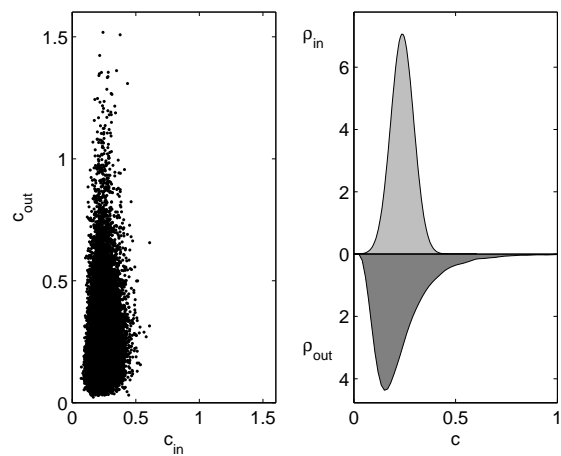}
\caption{{\rm For the EEG data the distributions of  $c_{in}$ and
$c_{out}$ are depicted in a scatter plot (left) and in terms of
their distributions, obtained by smoothing spline estimation
(right).\label{fig6}}}\end{figure}

Now we turn to consider real electroencephalogram (EEG) data. We
used recording obtained at rest from 10 healthy subjects. During the
experiment, which lasted for 15 min, the subjects were instructed to
relax and keep their eyes closed. To avoid drowsiness, every minute the subjects were
asked to open their eyes for 5 s. EEG was measured with a standard
10-20 system consisting of 19 channels \cite{nolte}. Data were
analyzed using the linked mastoids reference, and are available from
\cite{website_nolte}.

For each subject we considered several epochs of 4 seconds in which the subjects kept their eyes closed. For each
epoch we computed multivariate Kernel Granger Causality
\cite{noiprl} using a linear kernel and a model order of $5$,
determined by leave-one-out cross-validation. We then pooled all the
values for information flow towards and from any electrode and
analyzed their distribution.

\begin{figure}[ht]
\includegraphics[width=8.5cm]{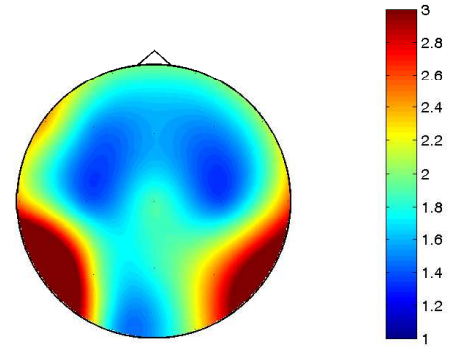}
\caption{{\rm The distribution on the scalp of  $R$ for EEG data. \label{fig7}}}\end{figure}

In figure (\ref{fig6}) we plot the incoming versus the outgoing
values of the information flow, as well as the distributions of the
two quantities: the incoming information seems exponentially
distributed whilst the outgoing information shows a fat tail. These
results suggest that overall brain effective connectivity networks
may also be considered in the light of the law of diminishing
marginal returns. More interestingly, this pattern is reproduced
locally but with a clear modulation:
 a topographic analysis has also been made considering the
distribution of incoming and outgoing causalities at each electrode.
In figure (\ref{fig7}) we depict the map of parameter $R$, thus
obtained, on the scalp;  the law of diminishing marginal returns
seems to affect mostly the temporal regions. This well defined
pattern suggests a functional role for the distributions.

It is worthwhile to mention that there are other measures of
directed brain connectivity, such as Directed Transfer Function,
Partial Directed Coherence and Phase Slope Index, which are not
directly related to information transfer. In particular, for DTF and
PDC, defined in terms of physiological bands, the interpretation in terms of information flow is still
debated \cite{takanashi}. On the other hand we verified
that a significant discrepancy between the distributions of incoming
and outgoing connectivities holds also for these methods.
Furthermore, bivariate measures do not display this asymmetry of the
distributions of $c_{in}$ and $c_{out}$: this is not surprising,
indeed it is well known that bivariate causality also account for
indirect interactions, see e.g. \cite{noipre}. Here we limited
ourselves to linear information flow; the amount of nonlinear
information transmission and its functional roles are not clear
\cite{noineuroimage}. It will be interesting to investigate these
phenomena also in the nonlinear case.

Summarizing, we have pointed out that the pattern of information
flows among variables of a complex system is the result of the
interplay between the topology of the underlying network and the
capacity of nodes to handle the incoming information. By means of
two simple toy models, we have shown that they may exhibit the law
of diminishing marginal returns for a suitable choice of parameters.
The analysis of a real EEG data-set has shown that similar patterns
exist for brain signals and could have a specific functional role.

\end{document}